\documentstyle[PASJadd]{PASJ95}
\draft

\newcommand{\vol}{}
\newcommand{\no}{}
\newcommand{\lett}{}
\newcommand{\spage}{}
\newcommand{\rdate}{2000 November 30}
\newcommand{\adate}{2001 February 20}

\begin{document}

\title{Tidal truncation of gas replenishment and global suppression of
galactic star formation in distant clusters}

\author{Kenji  {\sc Bekki} and Warrick J.  {\sc Couch}  \\
{\it 
School of Physics, University of New South Wales, Sydney 2052, Australia}\\
{bekki@bat.phys.unsw.edu.au, wjc@bat.phys.unsw.edu.au}  
\\[6pt]
and 
\\[6pt]
Yasuhiro  {\sc Shioya}\\
{\it Astronomical Institute,
Tohoku University, Sendai 980-8578} \\
{shioya@astro.tohoku.ac.jp}  
\\[6pt]
}

\abst{

Recent spectroscopic observations of galaxies in distant clusters
have revealed that the rate of star formation in star-forming 
galaxies is significantly suppressed with respect to their counterparts
in the field at a similar redshift. It is, however, highly uncertain 
which  physical processes are responsible for this suppression. We
present the results of a numerical investigation of how the global tidal
field of a cluster dynamically influences the reservoir of halo gas 
surrounding a disk galaxy as it falls into the cluster from the
surrounding field. 
We find that the tidal field of the cluster efficiently removes the halo 
gas from the galaxy, thereby halting its accretion onto the disk,  and thus
the fueling of star formation within. This effectively truncates the 
galaxy's star formation. We also find that this tidal truncation 
does not depend very strongly on the orbit of the disk
 with respect to the center of the cluster.
 These results suggest that the global
tidal field of clusters is capable of causing a widespread and uniform
suppression of star formation in galaxies accreted by the cluster. 
In light of these results, we discuss the importance of this tidal 
truncation of the gas supply in the  formation of S0 galaxies in
clusters.

}

\kword{
galaxies: clusters --- galaxies: star formation --- galaxies: ISM ---
galaxies: formation --- galaxies: interactions
 }

\maketitle
\thispagestyle{headings}

\section
{Introduction}

An impressive number of spectroscopic and photometric  results
have been accumulated which reveal that the star-formation histories of
galaxies in clusters are significantly different from  those in the field
(e.g., Dressler et al. 1985; Balogh et al. 1997; Couch et al. 1994;
Dressler et al. 1999; Poggianti et al. 1999). It is, however, still highly
uncertain which  physical processes specific to cluster environments govern
the observed difference. Starbursts triggered by some cluster-related
mechanism and the subsequent rapid consumption of gas are suggested to be
central to explaining the emission and absorption signatures seen
in cluster galaxy spectra (e.g., [$\rm O_{\rm II}$]$\lambda$3727 emission and strong
H$\delta$ absorption; Dressler,  Gunn 1983; Couch,  Sharples 1987; 
Barger et al. 1996; Poggianti et al. 1999). On the other hand, some
authors have claimed that if star formation is abruptly truncated when
galaxies are accreted by clusters, this is enough to explain the 
mean strengths of emission lines and any gradients with cluster-centric
radius (e.g., Newberry et al. 1990; Abraham et al. 1996; 
Balogh et al. 2000). Furthermore, Balogh et al. (2000) proposed 
an even `milder' scenario whereby star formation in cluster galaxies has
undergone just a gradual decline owing to the absence of halo gas that is
required to fuel star formation in galaxy disks. It remains unclear
which of the above three scenarios --- starbursts, abrupt truncation of
star formation, or gradual decline of star formation --- most accurately
describes the recent ($z<1$) star-formation history of the majority of
cluster galaxies. 

One of the difficulties in determining unambiguously the global star-formation
histories of cluster galaxy populations is that dust extinction 
greatly affects the spectroscopic properties of star-forming galaxies 
(e.g., Poggianti et al. 1999). The level of on-going and recent star-formation
activity inferred from [$\rm O_{\rm II}$] emission and H$\delta$ absorption
can be significantly distorted by dust (Smail et al. 1999; Poggianti, Wu 2000;
Bekki et al. in preparation).
In contrast to [$\rm O_{\rm II}$], the H$\alpha$ emission line is less 
affected by dust extinction (Kennicutt 1998),  and thus at least comes closer to
quantifying the rate of on-going star formation. 
Accordingly, Couch et al. (2000) have embarked on a wide-field ($\sim$ 4
$h^{-1}_{50}$ Mpc) H$\alpha$ survey of galaxies in 3 optically selected
clusters at {\it z}  $\sim$ 0.3. Their initial results show that star
formation within the members of one of these clusters ({\it AC}114) is strongly
and uniformly suppressed; within the region studied, at most only 10\% of
the cluster members are H$\alpha$-emitters, and their inferred star-formation 
rate is no more than $\sim 4\,M_{\odot}$\,yr$^{-1}$. Similarly, Balogh
and Morris (2000) found that only $\sim$ 12.5\% of the members within
the virial radius of Abell 2390 ($z = 0.23$) showed H$\alpha$ emission
with a rest-frame equivalent width greater than 50 \AA, thus arguing that
this is consistent with the `gradual decline' model of star formation.
Theoretical simulations by Balogh et al. (2000) have also shown that
if a galaxy's star formation undergoes a gradual decline after it has
been accreted by a cluster, the observed radial gradients in optical
colors and [$\rm O_{\rm II}$] emission line strength can be clearly explained.
Kodama and  Bower (2001) have also demonstrated that the gradual
decline of star formation is consistent with the distribution of galaxies in 
the color-magnitude diagrams for clusters in the range $0<z<0.4$. 
If the gradual decline of star formation is really the most promising  
scenario for the history of  star formation  in the cluster environment, a key
issue is which  physical processes cause such a decline. 

The purpose of this letter is to discuss our numerical investigation
of how the cluster 
environment can lead to a gradual decline in the star-formation rate of
its constituent members. In particular, our numerical simulations
have focused  on the disruption, by the global cluster tidal field, of the 
gas that is accreted onto galaxy disks from a surrounding gaseous halo,
and which fuels their star formation --- an effect first suggested by 
Larson, Tinsley, and   Caldwell (1980, hereafter LTC). Because the influence
of the cluster tidal field is so widespread, this would seem to be a natural 
mechanism for producing a significant suppression of star formation 
over large regions of the cluster. 
Furthermore, because the starvation of disk galaxies of their gas supply is highly
germane to the formation of S0 galaxies (see also LTC),  we
critically examine the cluster tidal truncation scenario in this context. 

The plan of this paper is as follows: In the next section we briefly
review our model and the numerical techniques and procedures used. The
results of our modeling are then presented in section 3. Finally, we
discuss our results and draw our conclusions in section 4.

\section
{Model}

The replenishment of interstellar gas due to sporadic and/or continuous
gas infall and acquisition from external environments (e.g., a gaseous
halo) has generally been considered to play a vital role in maintaining
star formation within disk galaxies up to the present day (LTC). 
We have investigated, numerically, the  effect which the global tidal field of a
rich cluster has on this important gas replenishment process in the
case of a disk galaxy that has been accreted by the cluster from the
surrounding field.

We have modeled the structure of a cluster  using the universal density profile
predicted by the standard cold dark-matter cosmogony (Navarro et al.
1996). The total cluster mass, represented by $M_{\rm cl}$,
the scale (core) radius ($r_{\rm c}$), and the virial radius ($r_{\rm v}$)
were  taken to be  $2.0 \times 10^{14}$ $M_{\odot}$, 127 kpc, and  1.16
Mpc, respectively, which are reasonable and realistic values for rich
clusters of galaxies such as Coma. 
The orbit of a disk galaxy accreted onto the cluster was assumed to
be influenced by an  $external$ $fixed$ gravitational potential resulting from 
the above-mentioned
cluster structure (i.e., the cluster potential is not ``live'' and thus not represented
by collisionless particles in the present simulations). 
Here, the disk galaxy was represented by 
Fall and  Efstathiou's (1980) bulgeless, purely stellar disk model. 
The initial ratio of the dark matter halo mass to the mass of disk stars
was taken to be 4:1. Values of 6.0 $\times$ $10^{10}$ $ \rm M_{\odot}$
and 17.5 kpc were  adopted for the disk mass ($M_{\rm d}$) and size ($R_{\rm
d}$), respectively.  The velocity and time are measured in units of 
$V_{\rm d}$ = $ (GM_{\rm d}/R_{\rm d})^{1/2}$  (1.21 $\times$ $10^{2}$
km\,s$^{-1}$) and $t_{\rm dyn}$ = $(R_{\rm d}^{3}/GM_{\rm d})^{1/2}$ (1.41
$\times$ $10^{8}$ yr), respectively, where $G$ is the gravitational
constant, which was  assumed to be 1.0 in the present study.
The exponential disk scale length and the vertical scale height 
are 3.5 \,kpc and 500 pc, respectively. 
The rotation curve became nearly flat at 0.35 radius with the maximum
velocity of $v_{\rm m}$ = 1.8 in our units.
We used the density profile of the dark matter halo described by
Fall and  Efstathiou (1980).
In addition to the disk's rotational velocity, initial radial and azimuthal 
velocity dispersions are also included in its motion according
to the epicyclic theory with Toomre's stability parameter $Q$ 
(Toomre 1964) set equal to 1.2.

We assumed  that the galaxy has a reservoir of halo gas from which gas
can be accreted onto the disk. It was assumed that the halo gas was 
uniformly and spherically distributed with a cut-off radius of $20R_{\rm
d}$ (0.35 Mpc). The halo gas, with a total mass of $0.1M_{\rm d}$, was 
represented by particles, each of which rotates around the disk with a
rotational velocity, $V_{\rm hg}$, satisfying the relation $V_{\rm
hg}R_{\rm hg}$ = $V_{\rm d}R_{\rm d}$, where $R_{\rm hg}$ is the initial
position of each halo particle. By adopting this relation, the specific
angular momentum of the halo gas is assumed not to be so remarkably
different from that of the disk,  accordingly, the halo gas is able to
fall onto the stellar disk region in  isolated evolution. A halo
particle is considered to have been accreted onto the disk
if it is within a vertical radius (z) of 500 pc. In our halo reservoir 
model, the accretion rate for the isolated evolution of the disk was 
estimated to be of the order of $\sim$ 1 $M_{\odot}$ ${\rm yr}^{-1}$ (for
$T>$ 4\,Gyr), which is consistent with the value required for maintaining
active star formation in disk galaxies (LTC; Blitz et al. 1999). 
Although the present numerical results depend on the initial size of 
the halo reservoir, we describe here only the results of the 
20\,$R_{\rm d}$ model, since its behaviour is very typical of the 
effects that the cluster tidal field has on such halo gas reservoirs. 
The adopted initial size of the halo reservoir is appreciably smaller than 
the observationally suggested mean distance ($\sim$ 1 Mpc from the center
of the Galaxy) of the High-Velocity Clouds (HVC) that have been
demonstrated to be halo gas reservoirs of the Galaxy (Blitz et al. 1999). 

Using this model, we investigated the dynamical evolution of the halo gas
reservoir for a variety of different orbits described by the disk 
galaxy upon its entry into the cluster. We present here the
results of three representative models. Figure 1 shows the orbital
evolution of the disk with three different initial positions in the
cluster. For example, the disk with initial $x$ and $y$ positions equal to
1.73 and 1.15 Mpc, respectively, passes by its pericenter ($x$ = 0.49 and
$y$ = 0.42 Mpc) at $T$ = 2.06 Gyr in orbit 1. 

The total number used in the simulations is 
10000 for the dark-matter halo of the disk, 12000 for the disk
galaxy, and 10000 for the halo gas reservoir. In order to demonstrate
more clearly the importance of cluster tidal effects 
on the halo gas reservoir, we treated the halo particles as
collisionless ones. Accordingly, the evolution of the halo `gas' reservoir
was determined only by purely gravitational evolution (dissipative effects
do not play a role at all in the evolution). Here, the softening length was 
1.09\,kpc for gravitational interactions  between the disk stars,
the halo gas components, and the dark matter halo surrounding the disk. 
We used these three different gravitational softening lengths to
investigate how the global tidal field of the cluster affects the local
dynamical evolution of the halo gas reservoir in an admittedly
self-consistent manner. 
A  direct summation method was used in a force calculation 
on a GPAPE board (Sugimoto et al. 1990).
 The energy and  angular momentum were  conserved within 1\% accuracy
in a test collisionless GRAPE simulation
of an isolated disk model without halo-gas replenishment.
\section
{Results}

As is shown in figure 2, tidal effect of the cluster has a strong
dynamical impact on the halo gas reservoir of the disk,  and consequently
greatly changes the mass distribution of the reservoir. As the disk enters
the cluster region ($T=1.70$\,Gyr), the halo gas begins to be distorted
owing to the tidal field of the cluster. After the galaxy passes the
pericenter (640 kpc corresponding to $5\,r_{\rm c}$) of the orbit
($T=2.23$\,Gyr), the cluster's global gravitational field dramatically
transforms the initially spherical halo into a rather elongated bar-like
structure ($T=2.83$\,Gyr). As the disk leaves the inner region of the
cluster ($T=3.95$\,Gyr), the gas reservoir is completely destroyed and
dispersed into the intra-cluster regions so that the mean gas density of
the halo becomes rather low compared with the initial value.  
These results demonstrate that the outer halo gas reservoir, which is
critically important for continuously supplying fresh gas to the disk, is  
fragile and thus susceptible to the tidal field of the cluster.
We thus  confirm here that stripping of the halo gas reservoir is
an important physical effect in cluster environments.

As a natural result of this, the accretion rate of halo gas onto the
disk is considerably reduced, both for $R$ $\le$ $R_{\rm d}$ and for 
$R_{\rm d}$ $<$$R$ $<$ 2$R_{\rm d}$ (see figure 3). When the galaxy 
passes the pericenter ($\sim$ 640 kpc) at $T=2.23$\,Gyr, the accretion
rate is essentially truncated, particularly, for $R$ $\le$ $R_{\rm d}$.
The mean accretion rate for $R$ $\le$ $R_{\rm d}$ is 0.16 $M_{\odot}$
${\rm yr}^{-1}$ for $T\le 2.23$\,Gyr and  0.42\,$M_{\odot}$ ${\rm
yr}^{-1}$ for $T>2.23$\,Gyr 
in the isolated disk model.
On the other hand, it is 0.05 $M_{\odot}$ ${\rm yr}^{-1}$
for $T\le 2.23$\,Gyr and $\sim$ 0 $M_{\odot}$ ${\rm yr}^{-1}$ for 
$T>2.23$\,Gyr in the galaxy with orbit 1.  
Although the galaxy does not pass through the cluster core
region ($R$ $<$ $r_{\rm c}$), the accretion rate is drastically reduced.
The derived effect is very different from other physical effects 
thought to operate in clusters,  such as ram pressure stripping (Dressler,
Gunn 1983; Farouki,  Shapiro 1980; Abadi et al. 1999) and the tidal effects of cluster
cores (Byrd, Valtonen 1990). Here, the latter two effects are important
only for galaxies passing through the cluster core. This is in contrast
to the results obtained here, which show that even if disk galaxies are
well outside the central region (even outside the virial radius), star
formation in disk galaxies is greatly suppressed due to  tidal
truncation of the gas-replenishment process.

Figure 4 demonstrates that irrespective of the orbit that the disk
galaxy takes through the cluster, the total mass accreted onto the disk
is drastically decreased owing to tidal stripping of the halo gas. The
ratio of accreted mass in the cluster model to that in the isolated one
is estimated to be typically 0.07 (0.20) for $R$ $\le$ $R_{\rm d}$
($R_{\rm d}$ $<$ $R$ $<$ $2 R_{\rm d}$). These results imply that
as a result of these disk galaxies being `starved' of their gas supply
once they are accreted by the cluster, their star formation will be
substantially curtailed within a few Gyr, and remain so irrespective of their location
within the cluster. 
Furthermore,  figure 4 combined with figure 2 suggests that even if a
galaxy is observed to be well outside the virial radius of a cluster, 
it  can show a rather small star-formation rate owing to this abrupt
truncation of its gas supply. 
Gas mass accretion is found to be strongly suppressed 
not only for the inner-disk regions ($R$ $\le$ $R_{\rm d}$),  but also for 
the outer ones ($R_{\rm d}$ $<$ $R$ $<$ $2 R_{\rm d}$; see figure 4),
which means that the size of the disk does not grow after the disk 
enters the cluster environment. If we consider  that truncation of gas
replenishment can also transform late-type disk galaxies into S0s (LTC),
the present results suggest that the typical disk size of
S0s is appreciably smaller than that of field disk galaxies
because of truncation of gas infall onto the outer parts of their disks.

\section
{Discussion}

We have demonstrated that the global tidal field of a cluster can greatly
reduce the halo gas-accretion rate in a disk galaxy accreted by the
cluster. The implication of this is that the disk rapidly consumes
(typically in a few Gyr; LTC) the remaining gas and, as a result, 
shows a subsequent decline in star formation after this event.
Thus, our numerical results have firstly confirmed the
earlier suggestion by LTC that the star-formation can be significantly
influenced by tidal effects in clusters. 
It is reasonable to say that star formation rate on a disk 
with truncation of star formation gradually declines after
the tuncation owing to no gas replenishment from the halo region.

Cluster tidal fields were  previously 
suggested to drive dramatic morphological transformations of
galaxies (Byrd, Valtonen 1990; Moore et al. 1996) and to induce rapid gas
consumption of $disk$ gas (Bekki 1998b). The `long-term' and less drastic
effects of cluster tidal fields on $halo$ gas derived in this study are
in striking contrast to those other `short-term' and rather dramatic
effects. In combination, however, they play a vital role in both triggering
transient enhancements in the star-formation rate and in reducing 
the mean star-formation rate in galaxies over a few Gyr.
 
Tidal effects between galaxies (Moore et al. 1996) and ram pressure stripping
of halo gas (Bekki et al. in preparation), both of which were  not modelled
in the present study at all,  are also suggested to efficiently remove the 
halo gas reservoirs surrounding disk galaxies in clusters.
We,  accordingly,  suggest that the present study, investigating $only$
cluster tidal fields, gives only a partial account of how the 
cluster environment truncates gas replenishment, which leads to a
substantial reduction in the rate of star formation within its
galaxies.

Abrupt truncation of gas replenishment from halo regions
in disks is suggested to transform gas-rich spiral galaxies into S0s (LTC).
If this truncation is a dominant mechanism for S0 formation, the present study
would provide the following three implications concerning the  nature of S0 galaxies in clusters.
Firstly, the color gradients within S0s can be appreciably shallower than those 
in disk galaxies, if metallicity gradients within the disk are their main cause. 
The radial dependence of chemical evolution driven mainly by star formation 
is considered to be responsible for the metallicity gradients within disk galaxies 
(e.g., de Jong 1996). Accordingly, if truncation of gas infall onto a disk galaxy  
strongly suppresses the star formation, and thus the chemical evolution, the galaxy 
should show smaller color (metallicity) gradients after its transformation into 
an S0. Secondly, the disk size could be  smaller for an S0 that has been transformed
from a disk galaxy after its accretion by the cluster at an
earlier epoch (or higher redshift). The present numerical simulations
demonstrate that gas infall onto the outer part of a disk ($R_{\rm d}$ $<$ $R$ $<$ 
$2R_{\rm d}$) is strongly suppressed,  and thus any growth in disk size
is prevented. Consequently, the disk size of an S0 would be  determined by
the epoch when the progenitor spiral galaxy was accreted by the cluster.
Therefore, the disk size is smaller for an S0 formed (by truncation) at higher
redshift, if the higher redshift spiral galaxies have smaller disks.
Thirdly, S0s can be formed by the gas-truncation process even in low-density 
regions of clusters. This is consistent with the recent observational evidence 
of Fasano et al. (2000) that the transformation of spirals into S0s must 
have occurred efficiently even in low-density regions in intermediate
redshift clusters. 

Finally, we stress that S0s formed by the above $gradual$ transformation, 
and which do not involve any `starburst', would have suffered a different fate
to those that have been formed by the type of tidal interaction described
by Byrd and  Valtonen (1990), the galaxy `harrassment' process described by
Moore et al. (1996), and unequal-mass merging (e.g., Bekki 1998a), all of 
which inevitably trigger strong nuclear starbursts and a dramatic morphological change
within only several dynamical time scales. 
The stellar populations in the central bulge components 
can be different between S0s formed by the `gradual' truncation
and those by interaction/merging. 
In particular, the bulges of S0s formed by the former process
will have redder optical colors and weaker Balmer absorption lines
than those formed via the latter process.
Future imaging/spectroscopy at  high resolution
will be of much interest in discreminating between these two different
evolutionary scenarios.

\par
\vspace{1pc}\par
Y. S. thanks  the Japan Society for Promotion of Science (JSPS)
Research Fellowships for Young Scientist.

\section*{References}
\small

\re
Abadi, M. G., Moore, B., \& Bower, R. G. 1999, MNRAS, 308, 947

\re
Abraham, R. G., Smecker-Hane, T.-A., Huchings, J. B., Carlberg, R. G.,
Yee, H. K. C., Ellingson, E., Morris, S., Oke, J. B., \& Rigler, M.  1996, ApJ, 471, 694

\re
Balogh, M. L., Morris, S. L., Yee, H. K. C., Carlberg, R. G.,
\& Ellingson, E. 1997, APJ, 488, L75

\re
Balogh, M. L., Navarro, J. F., \& Morris, S. L. 2000, (astro-ph/0004078)

\re
Balogh, M. L., \& Morris, S. L. 2000, ApJ, in press  (astro-ph/0007111)

\re
Barger, A. J., Aragon-Salamanca, A., Ellis, R. S., Couch, W. J.,
Smail, I., \& Sharples, R. M. 1996, MNRAS, 279, 1

\re
Bekki, K.  1998a, ApJ, 502, L133

\re
Bekki, K.  1998b, ApJ, 510, L15

\re
Bekki, K.  Couch, W. J., Shioya, Y. 2000, in preparation


\re
Blitz, L., Spergel, D. N., Teuben, P. J., 
Hartmann, D., \& Burton, W. B. 1999, ApJ, 514, 818

\re
Byrd, G., \& Valtonen, M., 1990, ApJ, 350, 89 

\re
Couch, W. J., \& Sharples, R. M., 1987, MNRAS, 229, 423

\re
Couch, W. J., Ellis, R. S., Sharples, R. M., \&  Smail, I.
1994, ApJ, 430, 121

\re
Couch, W. J., Balogh, M. L., Bower, R. G., Smail, I., Glazebrook, K.,
\& Taylor, M. 2000,  ApJ , submitted  

\re
de Jong, R. S. 1996, A\&A, 313, 377

\re
Dressler, A., \& Gunn, J. E. 1983, ApJ, 270, 7


\re
Dressler, A., et al. 1997, ApJ, 490, 577 

\re
Dressler, A., Smail, I., Poggianti, B. M., Butcher, H., Couch, W. J.,
Ellis, R. S., \& Oemler, A., Jr. 1999, ApJS, 122, 51

\re
Fall, S. M., \& Efstathiou, G. 1980, MNRAS, 193, 189

\re
Fasano, G., Poggianti, B. M., Couch, W. J., Bettoni, D., Kjaergaad, P., \& Mariano, M.
2000, preprint (astro-ph/0008195)

\re
Farouki, R., \& Shapiro, S. L.
1980, ApJ,   241, 928

\re
Kennicutt, R. C., Jr. 1998, ARAA, 36, 189

\re
Kodama, T., \& Bower, R. G., 2001, MNRAS, in press

\re
Larson, R. B., Tinsley, B. M., \& Caldwell, C. N. 1980, ApJ, 237, 692 (LTC)

\re
Moore, B., Katz, N., Lake, G., Dressler, A., \& Oemler, A., Jr. 1996,
ApJ, 379, 613

\re
Navarro, J. F., Frenk, C. S., \& White, S. D. M.
1996, ApJ, 462, 563

\re
Newberry, M. V., Boroson, T. A., Kirshner, R. P., 1990, ApJ, 350, 585

\re
Poggianti, B. M., \& Wu, H. 2000, ApJ, 529, 157

\re
Poggianti, B. M., Smail, I., Dressler, A., Couch, W. J., Barger, J.,
Butcher, H., Ellis, R. S., \& Oemler, A., Jr. 1999, ApJ, 518, 576

\re
Smail, I., Morrison, G., Gray, M. E., Owen, F. N., Ivison, R. J.,
Kneib, J.-P., \& Ellis, R. S. 1999, ApJ, 525, 609

\re
Sugimoto, D., Chikada, Y., Makino, J., Ito, T., Ebisuzaki, T., \& 
Umemura, M. 1990, Nature, 345, 33

\re
Toomre, A. 1964, ApJ, 139, 1217

\label{last}

\newpage

\section*{Figure Captions}

\begin{fv}{1}{18pc}%
{

Orbital evolution of a disk galaxy accreted onto a cluster 
with the mass of 2 $\times$ $10^{12}$ $M_{\odot}$ for three differnt
initial positions of the galaxy. The projected initial $x$ and $y$ positions are 
1.73 Mpc (corresponding to 1.5 $r_{\rm v}$ where $r_{\rm v}$ is the virial radius
of the cluster) and 1.16 Mpc ($r_{\rm v}$), respectively, for orbit 1,
1.73 Mpc and 0.64 Mpc (5 $r_{\rm c}$ where $r_{\rm c}$
is the cluster core radius) for orbit 2, and 1.73 Mpc and 0.13 Mpc ($r_{\rm c}$)
for orbit 3. The
initial  velocity of the disk is $-8.6$  $\times$ $10^2$ km ${\rm s}^{-1}$ for
the $x$ direction and 0 km ${\rm s}^{-1}$ for $y$ in all  three models. 
Several positions of the disk are indicated by crosses along
each of the three orbits for $T$ = 1.70, 2.23, 2.83, 3.95, and 5.66 Gyr. The
{\it small} and {\it large dotted} circles represent the core (or scale) 
radius and the virial radius, respectively. The frame measures 7.7 Mpc on a side.
}    
\end{fv}

\begin{fv}{2}{18pc}%
{

Mass distribution of halo gas at each time ($T$) 
projected onto the $x$--$y$ plane for the disk galaxy that follows orbit 1.
Each frame measures 1.75 Mpc on a side. Here,  the center of mass of the halo gas
is located at ($x$,$y$) = (0,0) for all time-steps,  so that the 
transformation of the halo gas reservoir can be more clearly seen. 
Note that the cluster tidal field can completely destroy the disk's halo gas reservoir
and,  consequently,  strip the gas from the disk.
}    
\end{fv}

\begin{fv}{3}{18pc}%
{

Time evolution of gas accretion rates (in units of $M_{\odot}$ ${\rm yr}^{-1}$) 
onto the region with  $R$ $\le$ $R_{\rm d}$ (upper) and onto that with 
$R_{\rm d}$ $<$ $R$ $<$ $2R_{\rm d}$ for an isolated disk (represented by 
the {\it dotted} line) and the disk which passes through the cluster 
on orbit 1 (a {\it solid} line). Note that at $T$ $>$  2.5 Gyr, the gas accretion 
is abruptly truncated.
}    
\end{fv}

\begin{fv}{4}{18pc}%
{

Time evolution of the total gas mass accreted onto 
the inner disk region ($R$ $\le$ $R_{\rm d}$; {\it upper panel})
and that with $R_{\rm d}$ $<$ $R$ $<$ $2R_{\rm d}$ 
({\it lower panel}) for our three different orbit models (1,2,3).
Here, 
the mass is given in our units ($M_{\rm d}$ corresponding to
6.0 $\times$ $10^{10}$ $M_{\odot}$). For a  comparison, the results
of the isolated disk model are also given (see {\it solid lines}).
The results of the models for orbits 1, 2, and 3 are represented
by the {\it dotted, short-dashed}, and {\it long-dashed} lines, respectively.
}    
\end{fv}

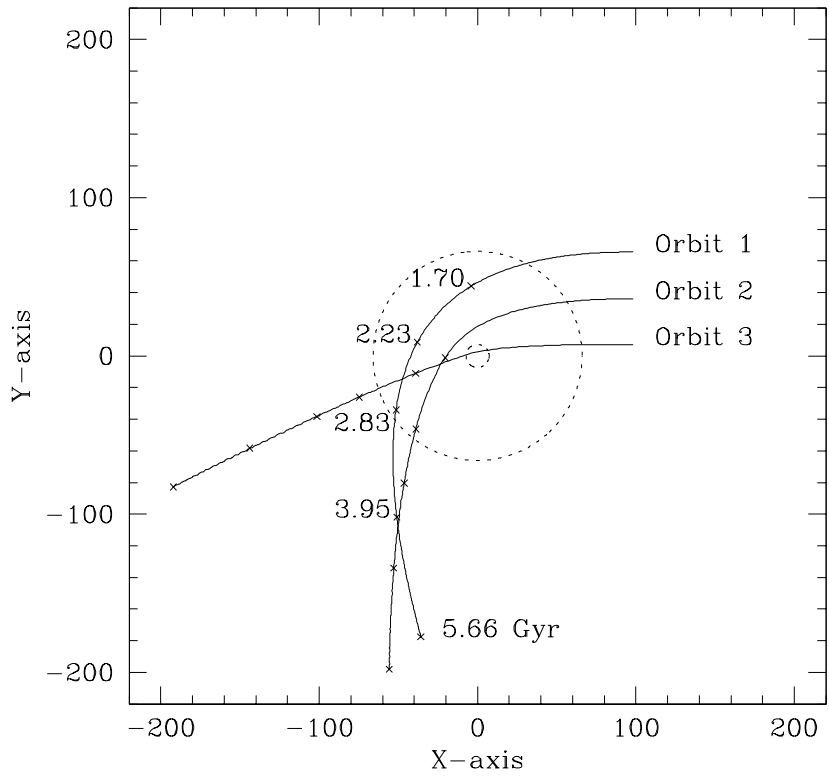
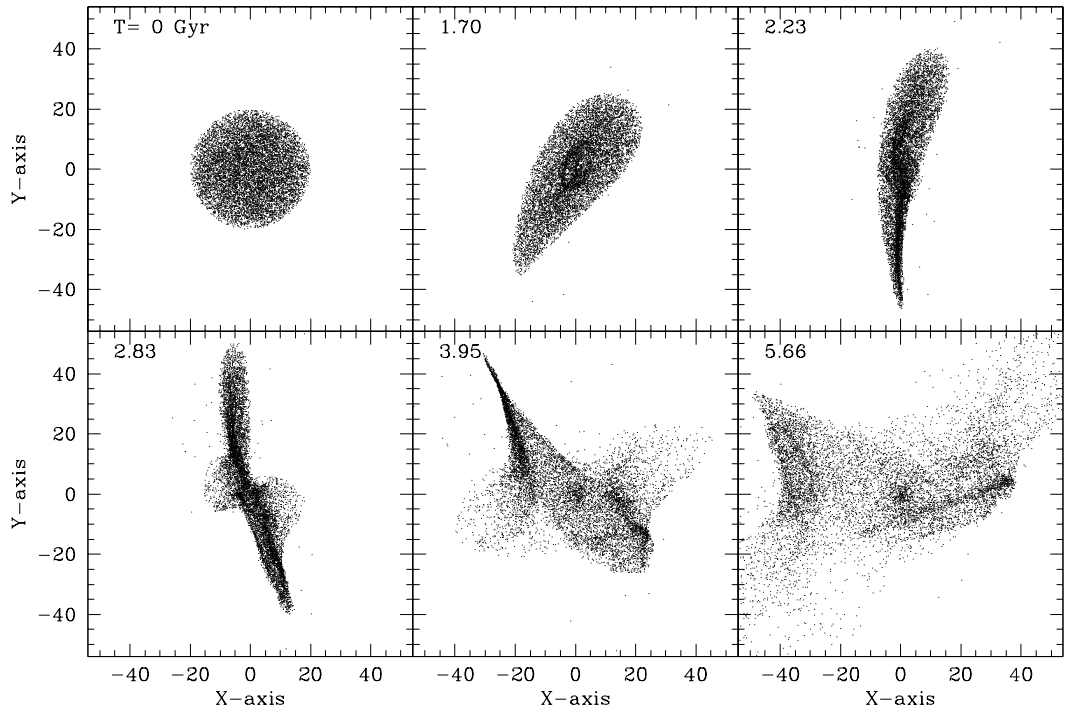
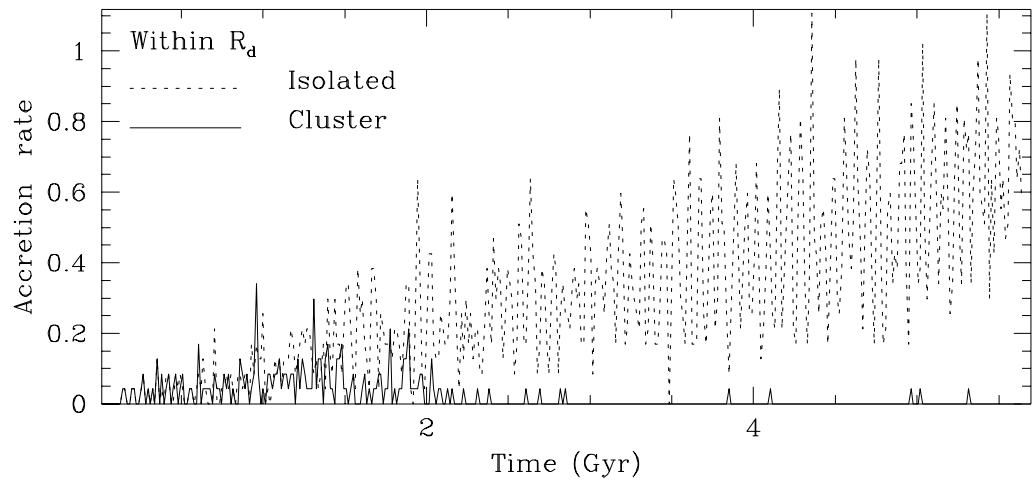
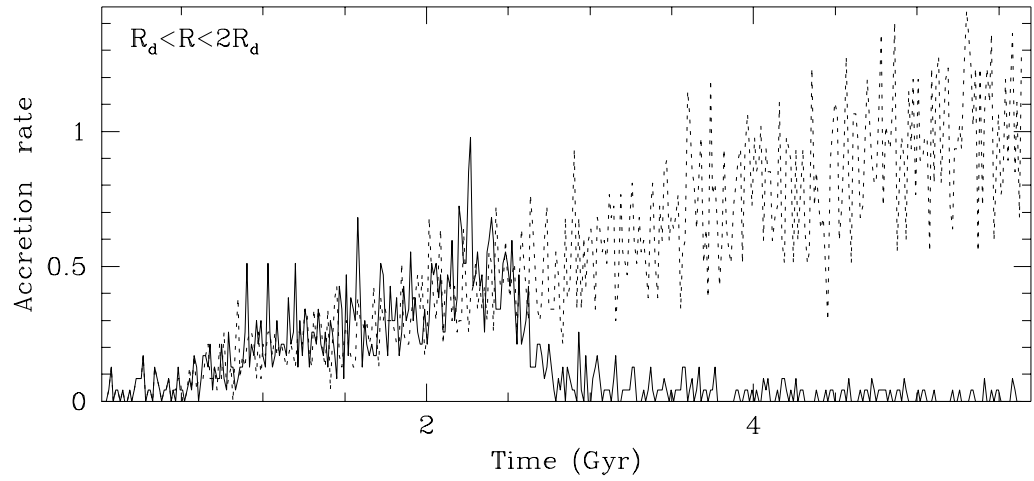
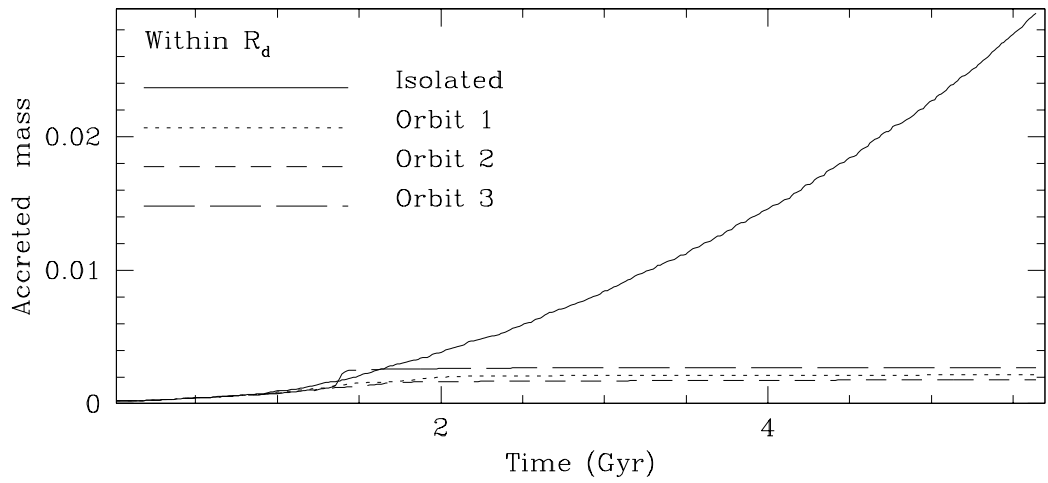
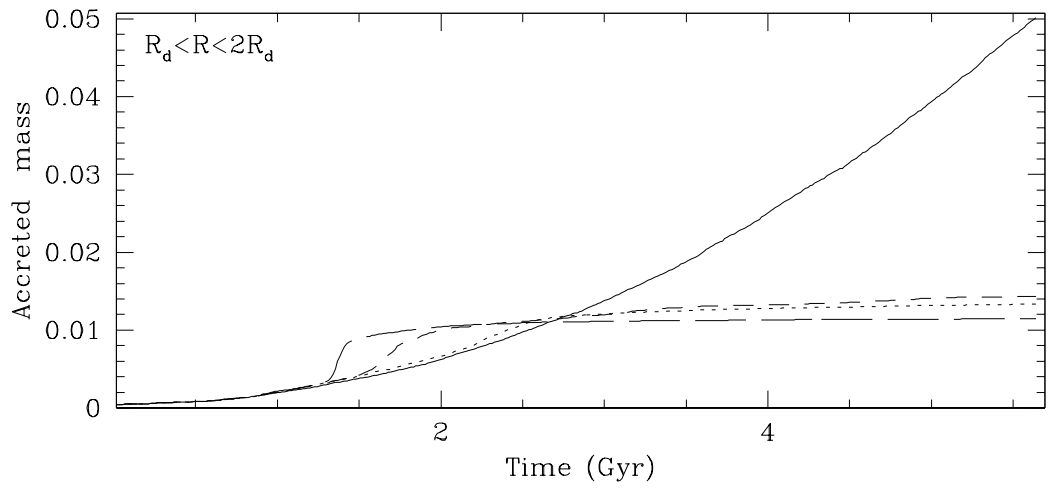
\end{document}